\documentclass{PoS}

\title{Deeply virtual Compton scattering on longitudinally polarized protons and neutrons at
CLAS}

\ShortTitle{Deeply virtual Compton scattering on longitudinally polarized protons and
neutrons at CLAS}

\author{\speaker{Silvia Niccolai}%
         \thanks{for the CLAS Collaboration}\\
        IPN-Orsay\\
        E-mail: \email{silvia@jlab.org}}

\abstract{This paper focuses on a measurement of deeply virtual Compton scattering (DVCS) performed at Jefferson Lab using a nearly-6-GeV polarized electron beam, two longitudinally polarized (via DNP) solid targets of protons ($NH_3$) and deuterons ($ND_3$) and the CEBAF Large Acceptance Spectrometer. 
Here, preliminary results for target-spin asymmetries and double (beam-target) asymmetries for proton DVCS, as well as a very preliminary extraction of beam-spin asymmetry for neutron DVCS, are presented and linked to Generalized Parton Distributions.}

\FullConference{Sixth International Conference on Quarks and Nuclear Physics\\
                 April 16-20, 2012\\
                 Ecole Polytechnique, Palaiseau,  Paris}
\begin{document}
\section{GDPs and DVCS}
The Generalized Parton Distributions (GPDs), first introduced nearly two decades ago, have emerged as a universal tool to describe hadrons, and nucleons in 
particular, in terms of their elementary constituents, the quarks and the 
gluons \cite{muller,ji}. The GPDs, which 
generalize the features of form factors and ordinary parton distributions, 
describe the correlations between partons in quantum states of different 
(or same) helicity, longitudinal momentum, and transverse position. 
They also can give access, via the Ji's sum rule \cite{ji}, to the 
contribution to the nucleon spin coming from the orbital angular momentum 
of the quarks.
At leading order and at twist 2 there are four different GPDs for the nucleon: $H$, $E$ (the two spin-independent GPDs), $\tilde{H}$, $\tilde{E}$ (the two 
spin-dependent GPDs), and they can be measured in exclusive hard reactions.

\begin{figure}[h]
\begin{center}
{\includegraphics[scale=0.8]{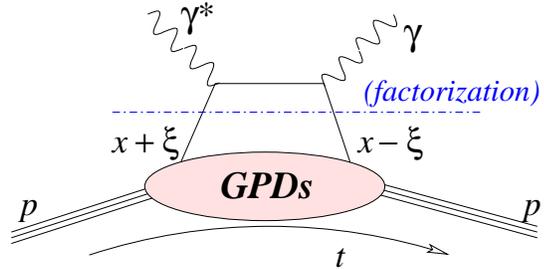}}
\caption{Handbag diagram for the DVCS reaction. $x$ is the average longitudinal momentum fraction of the active quark in the initial and final states, while $2\xi$ is their difference ($\xi\simeq x_B/(2-x_B)$, where $x_B$ is the Bjorken scaling variable). The third variable on which the GPDs depend is $t=(p'-p)^2$, the squared four-momentum transferred to the target. }
\label{handbag}
    \end{center}
\end{figure}

Deeply Virtual Compton scattering (DVCS) ($eN \to e'N'\gamma$) is the simplest process to access GPDs of the nucleon ($N$). In the Bjorken regime (high $\gamma^*$ virtuality $Q^2$ and small 
squared momentum transferred to the nucleon $t$) and at leading twist, 
this mechanism corresponds to the absorption 
of a virtual photon by a quark carrying the longitudinal 
momentum fraction $x+\xi$. The struck quark emits a real photon and 
goes back into the nucleon with longitudinal momentum fraction 
$x-\xi$. The amplitude for DVCS is factorized \cite{ji} into a 
hard-scattering part (exactly calculable in pQCD) and a non-perturbative 
part, representing the soft structure of the nucleon, parametrized by 
the GPDs, which will depend on the three kinematic variables $x$, 
$\xi$, and $t$ (Fig.~\ref{handbag}). 

The DVCS amplitude interferes with the amplitude for Bethe-Heitler (BH), the process where the real photon is emitted either by the incoming or the scattered electron. The DVCS-BH interference gives rise to spin asymmetries, which can be connected to combinations of GPDs. For instance, referring to proton DVCS, the beam-spin asymmetry (BSA) is particularly sensitive to the imaginary part of the GPD $H$, while the contribution of the other GPDs are expected to be negligible \cite{vgg}. The target-spin asymmetry (TSA), for a longitudinally polarized proton target, can instead give access to a combination of the imaginary parts of the GPDs $H$ and $\tilde{H}$. The double beam-target asymmetry is a way to access the real part of the GPDs $H$ and $\tilde{H}$. 
Information about the flavor decomposition of the GPDs via DVCS requires measurements with both proton and neutron targets. Depending on the target nucleon and on the polarization observable extracted, different sensitivity to the four GPDs ($H$, $E$, $\tilde{H}$, $\tilde{E}$) for each quark flavor ($u$, $d$) can be exploited. 

After the first pioneering observation of $\sin\phi$ behavior --- 
hinting to handbag dominance --- on the DVCS beam-spin asymmetry 
obtained from non-dedicated data \cite{stepan} taken with the CLAS 
detector \cite{clas}, various DVCS- and GPD-focused experiments 
have been performed at Jefferson Lab. As of today, the results of DVCS-dedicated experiments performed in 
Hall A (polarized and unpolarized cross sections \cite{carlos}) and 
in Hall B with CLAS (BSA \cite{fx} and TSA with longitudinally polarized 
target \cite{shifeng}) suggest that the handbag mechanism dominates 
already at relatively low $Q^2$ ($\sim 2$ GeV$^2$). More recent CLAS data taken at 6 GeV with unpolarized 
and longitudinally polarized proton and deuteron targets 
are under analysis, and new results for proton-DVCS cross sections \cite{hs}, TSAs \cite{erin}, double-spin (beam-target) asymmetry \cite{gary} and BSA for neutron DVCS are currently being produced. The last three observables are the focus of this paper. 

\section{The CLAS eg1-dvcs experiment}
The data presented here were taken at Jefferson Lab during the eg1-dvcs experiment, that ran from the spring to the early fall of 2009. The CEBAF polarized electron beam (of energies between 4.735 and 5.967 GeV) was delivered to the Hall B, where it impinged on a solid dynamically polarized target \cite{chris}. $NH_3$ was used as target material for two thirds of the duration of the experiment (proton-DVCS analysis), while $ND_3$ was adopted for the last third (neutron-DVCS analysis). The scattered electron and recoil nucleon were detected in CLAS, while the majority of the DVCS/BH photons were detected by the Inner Calorimeter, an array of 424 lead-tungstate crystals readout by avalanche photodiodes, which completed the CLAS acceptance for photons for polar angles between $4.5^o$ and $15^o$. 
Once all the three DVCS final-state particles ($ep\gamma$ or $en\gamma(p)$) were identified and their momenta and angles measured, constraints such as cuts on missing energy and mass could be applied, in order to ensure the exclusivity of the reaction by minimizing the background contamination --- the main source of background being the $ep\pi^0$ channel, where one of the two $\pi^0$-decay photons had escaped detection. 
The final event sample was finally divided according to the sign (positive or negative with respect to the beam direction) of the target polarization and of the beam helicity, and asymmetries were constructed (compatibly to the available statistics) as functions of the 4 relevant kinematic variables of DVCS ($Q²$, $x_B$, $-t$, and the angle $\phi$ formed by the leptonic and the hadronic planes). The covered kinematic ranges are: 1 GeV$^2 <Q^2<3.5$ GeV$^2$, $0.15<x_B<0.42$, 0.1 GeV$^2 <-t<0.6$ GeV$^2$. 

\subsection{Results on proton DVCS}
\begin{figure}[h]
\begin{center}
{\includegraphics[scale=0.18]{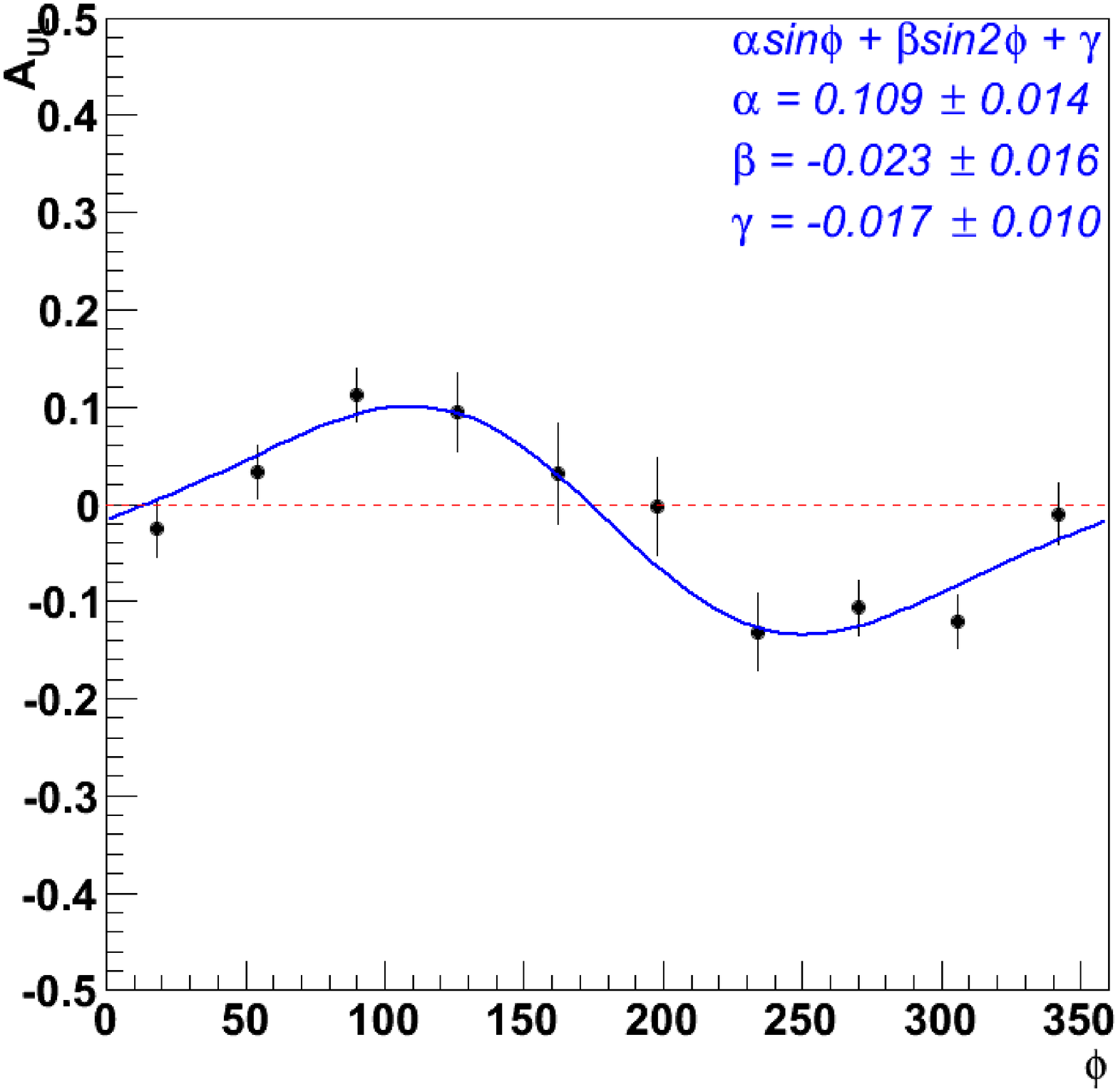}}
{\includegraphics[scale=0.7]{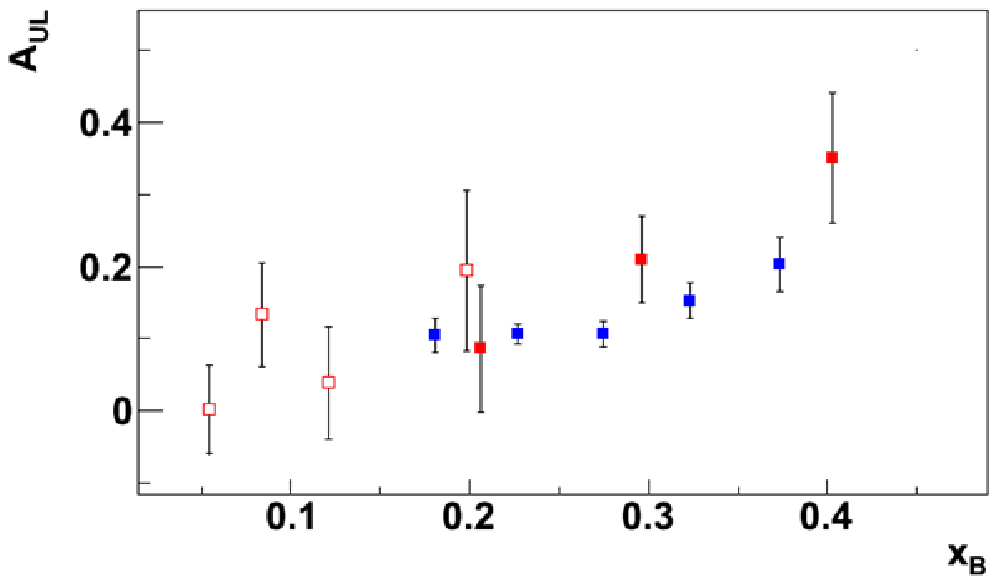}}
\caption{Left: target-spin asymmetry for proton DVCS, for one particular $x_B$ kinematic point ($x_B=0.23$), integrated over all values of $Q^2$ and $-t$, plotted as a function of $\phi$ and fitted with the $\alpha\sin\phi+\beta\sin 2\phi+\gamma$ function. Right: distribution of the $\alpha$ term as a function of $x_B$; the blue squares are the preliminary results from the CLAS eg1-dvcs data, the red squares are previous CLAS results from a non-dedicated experiment \cite{shifeng}, while the open red squares have been produced at HERMES \cite{HERMES}.}
\label{erin_plots}
    \end{center}
\end{figure}

\begin{figure}[h]
\begin{center}
{\includegraphics[scale=0.35]{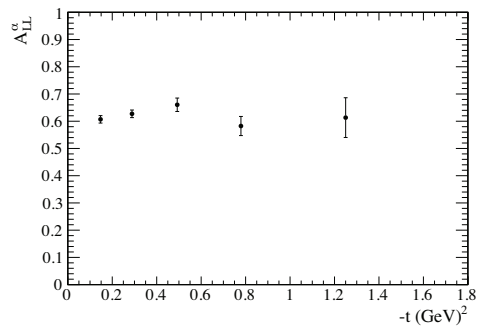}}
\caption{Preliminary results from eg1-dvcs on double beam-target asymmetry for proton DVCS: constant term $\alpha$, from fits of the $\phi$ distributions of the asymmetry with the function $\alpha+\beta\cos\phi$, plotted as a function of $-t$ (the average kinematics are $Q^2=2.39$, $x_B=0.31$).}
\label{gary_plot}
\end{center}
\end{figure}

Preliminary results for single target-spin asymmetries and double (beam-target) asymmetries are shown in Figs.~\ref{erin_plots} and~\ref{gary_plot}, respectively. No $\pi^0$ background subtraction has yet been applied at this stage for either of the two extracted observables. For each $x_B$ bin, the $\phi$ distribution of the target-spin asymmetry is fitted with the $\alpha\sin\phi+\beta\sin 2\phi+\gamma$ function. The dominance of the $\alpha$ term (left plot of Fig.~\ref{erin_plots}) confirms the validity fo the twist-2 approximation for the CLAS kinematics, as it was previously observed from the beam-spin asymmetries \cite{fx}. The right plot of Fig.~\ref{erin_plots} shows the $x_B$ dependence of the $\alpha$ term, comparing the eg1-dvcs preliminary results (blue points) with published results from HERMES (open red squares) \cite{HERMES} and the ones from a previous CLAS exploratory measurement \cite{shifeng}. The eg1-dvcs preliminary data are in good agreement with previos measurements, and improve of the statistical precision of the TSA by about a factor of 4. 
From preliminary fits to the double beam-target asymmetry (Fig.~\ref{gary_plot}) with the $\alpha+\beta\cos\phi$ function a strong contribution of the constant term $\alpha$ is observed, which hints to the dominance of the Bethe-Heitler mechanism over DVCS for this observable in our kinematics, as it was previously observed by the HERMES collaboration \cite{HERMES}. A higher statistical precision on this observable, such as the one that will be achieved in the proton-DVCS experiments planned for CLAS12 \cite{franck}, after the 12-GeV upgrade of the CEBAF accelerator, will therefore be necessary in order to have good sensitivity to DVCS - and hence to the GPDs. 

These data, combined with the published beam-spin asymmetries \cite{fx} and preliminary cross sections \cite{hs} from CLAS, are currently being used in a global-fitting procedure \cite{michel} of DVCS observables to extract the Compton Form Factors (CFFs) of the GPDs. For each GPD $F$, two CFFs are experimentally accessible via DVCS: 
\begin{eqnarray}
Re\mathcal{F} &=& P \int_0^{\infty}{dx[F(x,\xi,t)\pm F(-x,\xi,t)]C^{\mp} (x,\xi)}\\ 
Im\mathcal{F} &=& F(\xi,\xi,t)-F(-\xi,\xi,t)
\end{eqnarray}
where
\begin{eqnarray}
C^{\pm}(x,\xi)=\frac{1}{x-\xi}\pm\frac{1}{x-\xi},
\end{eqnarray}
and the two different combinations of signs correspond, respectively from top to bottom, to spin-independent ($H$, $E$) and spin-dependent ($\tilde{H}$, $\tilde{E}$) GPDs. 
Preliminary fits show that the inclusion of the eg1-dvcs single and double asymmetries is bringing an improvement of about a factor of 4 in the precision on the $Im{\tilde{\mathcal{H}}}$ Compton Form Factor. 

\subsection{Results on neutron DVCS}
\begin{figure}[h]
\begin{center}
{\includegraphics[scale=0.4]{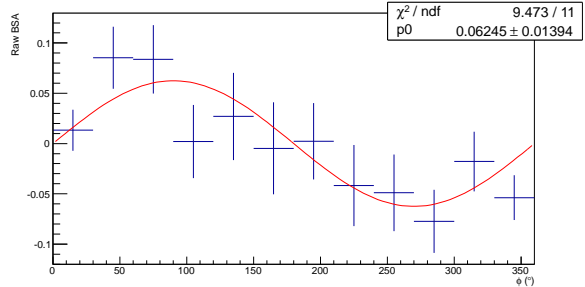}}
\caption{Very preliminary results from eg1-dvcs on beam-spin asymmetry for neutron DVCS as a function of $\phi$.}
\label{daria_plot}
\end{center}
\end{figure}

\begin{figure}[h]
\begin{center}
{\includegraphics[scale=0.3]{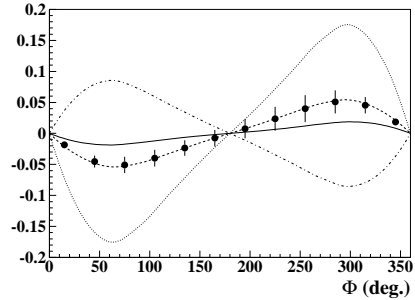}}
\caption{Projected beam-spin asymmetry for neutron DVCS, as a function of $\phi$, for $<Q^2>=2.75$ GeV$^2$, $<x_B>=0.225$, and $<-t>=0.35$  GeV$^2$. The points illustrate the error bars expected for the CLAS12 neutron-DVCS experiment. The curves are predictions by the VGG model \cite{vgg} for different values of the quarks' orbital momenta $J_u$ and $J_d$ that, in this model, parametrize the GPD $E$: $J_u=0.1$ and $J_d=0.1$ for the solid line, $J_u=0.3$ and $J_d=0.1$ for the dashed line, $J_u=0.3$ and $J_d=0.3$ for the dotted line, and  $J_u=0.3$ and $J_d=-0.1$ for the dashed-dotted line.}
\label{clas12_plot}
\end{center}
\end{figure}

Using the last part of the eg1-dvcs data set, that was taken using a polarized $ND_3$ target, a very preliminary raw (no background subtraction has been applied yet) beam-spin asymmetry of neutron DVCS was extracted \cite{daria} and it is shown in Fig.~\ref{daria_plot} as a function of $\phi$. This observable, which has previously been measured only in an exploratory Hall A experiment \cite{mazouz}, is crucial due to its unique sensitivity to the GPD $E$, which is otherwise only accessible through the experimentally-challenging transverse target-spin asymmetry for proton DVCS. Extracting the GPD $E$ is an unavoidable step towards the measurement of Ji's sum rule \cite{ji}, which, connecting the second moment in $x$ of the sum of $H$ and $E$ to the total angular momentum carried by the quarks, is a fundamental missing piece towards the unraveling of the ``spin crises''. 
The eg1-dvcs results for the neutron-DVCS BSA, albeit of poor statistical precision, display a non-negligible $\sin\phi$ dependence. This measurement is a very encouraging starting point for the neutron-DVCS experiment that is planned for CLAS12 \cite{ndvcs}, and that will provide an accurate mapping of the n-DVCS beam-spin asymmetry over a wide 4-dimensional (1 GeV$^2<Q^2<11$ GeV$^2$, $0.05<x_B<0.7$, 0 GeV$^2<-t<1.2$ GeV$^2$,$\phi$) phase space. Figure~\ref{clas12_plot} shows the expected sensitivity of the extracted asymmetry to different values of the GPD $E$ (parameterized, in the VGG model \cite{vgg}, by the quarks' total angular momenta $J_u$ and  $J_d$), for one of the 49 $Q^2$-$x_B$-$t$ bins for which we will extract the $\phi$ dependence (in 12 $\phi$ bins) of the BSA.  


\begin{thebibliography}{99}
\bibitem{muller} D. Muller {\it et al.}, Fortschr. Phys. 42, 101 (1994); A.V. Radyushkin, Phys. Rev. D {\bf 56}, 5524 (1997); M. Diehl, Eur. Phys. J. C {\bf 25}, 223 (2002).
\bibitem{ji} X.-D. Ji, Phys. Rev. Lett. {\bf 78}, 610 (1997).
\bibitem{vgg} M. Guidal, M.V. Polyakov, and M. Vanderhaeghen, Phys. Rev. 
D {\bf 72}, 054013 (2005).
\bibitem{stepan} S. Stepanyan {\it et al.}, Phys. Rev. Lett. {\bf 87}, 182002 (2001).
\bibitem{clas} B.A. Mecking {\it et al.}, Nucl. Instr. Meth. {\bf A503}, 
513 (2003).
\bibitem{carlos} C. Munoz Camacho {\it et al.}, Phys. Rev. Lett. {\bf 97}, 
262002 (2006).
\bibitem{fx} F.X. Girod {\it et al.}, Phys. Rev. Lett. {\bf 100}, 162002 
(2008).
\bibitem{shifeng} S. Chen {\it et al.}, Phys. Rev. Lett. {\bf 97}, 072002 
(2006).
\bibitem{hs} H.S. Jo, analysis underway.
\bibitem{erin} E. Seder, analysis in progress.
\bibitem{gary} G. Smith, analysis in progress.
\bibitem{chris} C. D. Keith, {\it et al.}, Nucl. Instr. Meth. {\bf A501}, 327 (2003).
\bibitem{HERMES} HERMES Collaboration, arXiv:1004.0177v1 [hep-ex].
\bibitem{franck} F. Sabati\'e {\it et al.}, JLab Experiment E12-06-119.
\bibitem{michel} M. Guidal, PoS ICHEP2010 (2010) 148.
\bibitem{daria} D. Sokhan, analysis in progress. 
\bibitem{mazouz} M. Mazouz, Phys. Rev. Lett. 99, 242501 (2007).
\bibitem{ndvcs} S. Niccolai {\it et al.}, JLab Experiment E12-11-003.
\end{thebibliography}
\end{document}